\documentclass[acmsmall]{acmart}
\AtBeginDocument{%
  \providecommand\BibTeX{{%
    Bib\TeX}}}

\setcopyright{acmlicensed}
\acmDOI{10.1145/3715750}
\acmYear{2025}
\acmJournal{PACMSE}
\acmVolume{2}
\acmNumber{FSE}
\acmArticle{FSE033}
\acmMonth{7}
\received{2024-09-13}
\received[accepted]{2025-01-14}

\usepackage[ruled,linesnumbered]{algorithm2e}
\usepackage{multirow}
\usepackage{listings}
\usepackage{booktabs}
\usepackage{multicol}
\usepackage{tabularx}
\usepackage{adjustbox}
\usepackage{subfig}
\usepackage{subcaption}
\usepackage{makecell}
\usepackage{enumitem}
\usepackage{calligra}
\usepackage{mdframed}
\usepackage{fontawesome}
\usepackage{amsmath}
\usepackage{amsfonts}
\usepackage{graphicx}
\usepackage{textcomp}
\usepackage{xcolor}
\usepackage{balance}
\usepackage{caption}
\usepackage{wrapfig}
\usepackage{colortbl}
\usepackage{forest}
\usepackage{array}
\usepackage[most]{tcolorbox}
\usepackage{pifont}

\def\BibTeX{{\rm B\kern-.05em{\sc i\kern-.025em b}\kern-.08em
    T\kern-.1667em\lower.7ex\hbox{E}\kern-.125emX}}

\lstdefinestyle{codeStyle}{
    basicstyle=\small\ttfamily,
    commentstyle=\color{green},
    keywordstyle=\color{blue},
    stringstyle=\color{blue},
    showstringspaces=false,
    breaklines=true,
    frame=lines,
    backgroundcolor=\color{white},
    captionpos=b,
    aboveskip=10pt,
    belowskip=10pt
}
\definecolor{lightgreen}{rgb}{0.9,1,0.9}
\definecolor{grayheader}{gray}{0.85}
\definecolor{codebackground}{RGB}{240,240,240} %

\newcommand{\appname}{{\sc UTRefactor}\xspace}

\begin{document}

\title{Automated Unit Test Refactoring}

\author{Yi Gao}
\email{gaoyi01@zju.edu.cn}
\orcid{https://orcid.org/0009-0000-2554-2381}
\affiliation{%
  \institution{The State Key Laboratory of Blockchain and Data Security, Zhejiang University}
  \city{Hangzhou}
  \country{China}
}

\author{Xing Hu}
\authornote{Corresponding Author}
\email{xinghu@zju.edu.cn}
\orcid{https://orcid.org/0000-0003-0093-3292}
\affiliation{%
  \institution{The State Key Laboratory of Blockchain and Data Security, Zhejiang University}
  \city{Hangzhou}
  \country{China}
}

\author{Xiaohu Yang}
\email{yangxh@zju.edu.cn}
\orcid{https://orcid.org/0000-0003-4111-4189}
\affiliation{%
  \institution{The State Key Laboratory of Blockchain and Data Security, Zhejiang University}
  \city{Hangzhou}
  \country{China}
}

\author{Xin Xia}
\email{xin.xia@acm.org}
\orcid{https://orcid.org/0000-0002-6302-3256}
\affiliation{%
  \institution{The State Key Laboratory of Blockchain and Data Security, Zhejiang University}
  \city{Hangzhou}
  \country{China}
}

\begin{abstract}
Test smells arise from poor design practices and insufficient domain knowledge, which can lower the quality of test code and make it harder to maintain and update.
Manually refactoring of test smells is time-consuming and error-prone, highlighting the necessity for automated approaches. 
Current rule-based refactoring methods often struggle in scenarios not covered by predefined rules and lack the flexibility needed to handle diverse cases effectively.
In this paper, we propose a novel approach called \appname, a context-enhanced, LLM-based framework for automatic test refactoring in Java projects. 
\appname extracts relevant context from test code and leverages an external knowledge base that includes test smell definitions, descriptions, and DSL-based refactoring rules.
By simulating the manual refactoring process through a chain-of-thought approach, \appname guides the LLM to eliminate test smells in a step-by-step process, ensuring both accuracy and consistency throughout the refactoring.
Additionally, we implement a checkpoint mechanism to facilitate comprehensive refactoring, particularly when multiple smells are present.
We evaluate \appname on 879 tests from six open-source Java projects, reducing the number of test smells from 2,375 to 265, achieving an 89\% reduction. 
\appname outperforms direct LLM-based refactoring methods by 61.82\% in smell elimination and significantly surpasses the performance of a rule-based test smell refactoring tool.
Our results demonstrate the effectiveness of \appname in enhancing test code quality while minimizing manual involvement.
\end{abstract}



\maketitle

\section{Introduction}
During the software testing process, test code often suffers from \textbf{test smells}, which originates from a lack of sufficient domain knowledge by software engineers and the adoption of poor design practices when writing test code. 
Recent studies~\cite{damasceno2022analyzing,pizzini2023sentinel} have shown that developers tend to prioritize production code over test code, which further contributes to the decline in test code quality.
From a software maintenance perspective, test smells in a software system can complicate the test code, making it harder to read, understand, and update.
This complexity can ultimately lead to a decline in software product quality and reduced developer productivity.
Test smells can be removed through test refactoring—a process of improving the internal structure of the code without altering the software's external behavior~\cite{de2022testaxe,soares2020refactoring,pizzini2023sentinel}. 
Numerous empirical studies~\cite{soares2020refactoring,pizzini2023sentinel,pizzini2022behavior,nagy2022co,damasceno2022analyzing,soares2022refactoring,kashiwa2021does} have highlighted the importance of test refactoring, and developers widely accept the refactored tests, and both developers and testers are increasingly recognizing the negative impact of test smells. 
They widely agree that test quality improves significantly once test smells are eliminated.
Additionally, many studies~\cite{peruma2020tsdetect,yang2024lost,aljedaani2021test,palomba2018automatic,pontillo2024machine,wang2021pynose,danphitsanuphan2012code,fernandes2022tempy,bodea2022pytest,virginio2020jnose} have proposed test smell detection tools across programming languages such as Java, Scala, and Python, while also evaluating the impact of test smells on the software development process.

However, manual refactoring of test code remains time-consuming, inefficient, and is prone to errors, underscoring the need for automated tools to address a broader range of test smells.
Although there is a clear demand for such tools, only a few open-source options exist for automatic test refactoring~\cite{de2022testaxe}. 
Those tools primarily handle basic and limited types of test smells, leaving many types unaddressed.
At present, no tool is capable of efficiently and automatically eliminating all types of test smells.

Given the remarkable capabilities of large language models (LLMs) in understanding, generating, and reviewing code, they might have the potential to play a key role in test refactoring tasks.
In this paper, we explore the potential of open-source LLMs to automate unit test refactoring and propose an approach for automatically eliminating test smells in software projects. However, several challenges must be addressed to achieve this goal:

\textbf{Challenge 1: How to guide the LLM to eliminate test smells in an expected way?}
Refactoring aims to optimize code structure without altering the original functionality and logic. 
However, LLMs may generate hallucinations during the refactoring process, producing random code segments that introduce syntax or semantic errors. 
Thus, the first challenge is how to direct the LLM to follow predefined steps for refactoring, ensuring it remains consistent and produces the intended results.

\textbf{Challenge 2: How to eliminate multiple test smells simultaneously?} 
Since there are many types of test smells, and multiple smells may exist within a single test method, different removal orders can lead to different refactoring outcomes. 
The challenge lies in ensuring that all identified test smells are accurately and comprehensively eliminated.

In this paper, we propose a context-enhanced, LLM-based automatic test refactoring approach for Java projects.
Our proposed approach, named \appname, first extracts the project's test code and related context information, such as the methods and classes under test.
We then build an external knowledge base that supports test smell elimination, which includes test smell definitions and descriptions, DSL-based refactoring rules, and other relevant contextual information. 
Next, we simulate the manual refactoring process typically followed by developers or testers. 
Using a chain-of-thought approach, we guide the LLM through understanding the test's intent, identifying the test smells, and following the DSL-defined refactoring steps to refactor the test code. 
Additionally, we design a checkpoint mechanism to ensure more thorough smell elimination, particularly when multiple test smells are present in a single test.

We evaluate \appname on six popular Java open-source projects collected from GitHub. 
In our experiments, we detect 879 tests with smells out of 9,149 tests and apply smell elimination refactoring to them. 
The results show that \appname reduces the number of test smells from 2,375 to 265, achieving a reduction rate of 89\%. 
This represents a 61.82\% improvement in performance compared to directly using an LLM for test refactoring.
The main contributions of this paper are as follows:
\begin{itemize}[leftmargin=*]
\item  We propose a novel approach that leverages code context information and integrates refactoring DSL rules to enhance the test refactoring capabilities of LLMs, enabling the automatic elimination of test smells and improving the quality of unit tests.
\item We develop \appname, a tool that assists developers and testers in automatically refactoring test code in Java projects. 
It supports various levels of granularity, including single tests, test files, and entire projects. 
\item We evaluate \appname on 879 tests across six open-source projects, reducing the number of test smells from 2,375 to 265, outperforming baselines in effectiveness.
\end{itemize}
\section{Motivation}
\label{sec:motivation}
\begin{figure}
\centerline{\includegraphics[width=\textwidth]{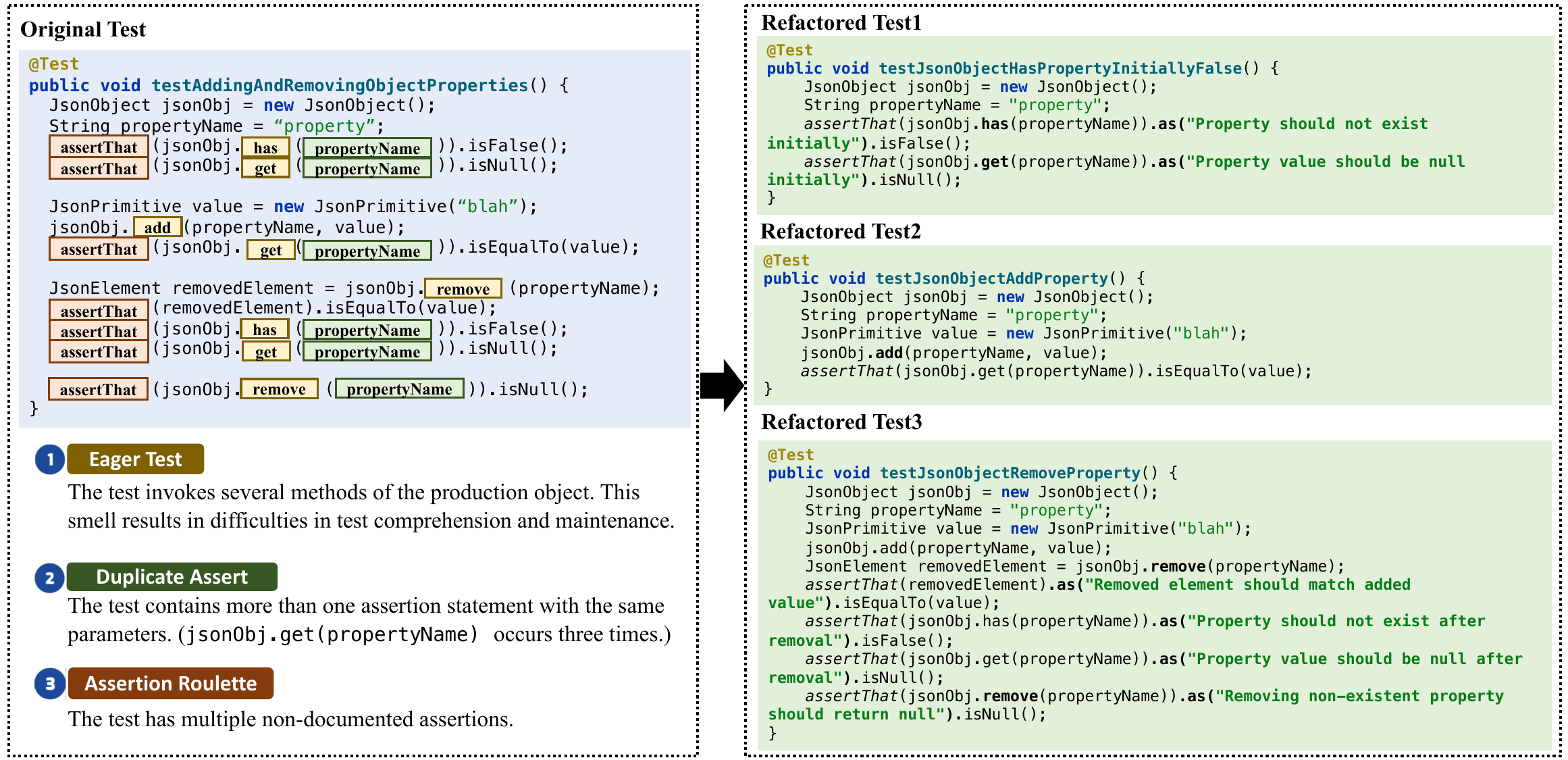}}
\caption{An example of test refactoring within the Gson.}
\label{fig:moti}
\end{figure}

Developers often prioritize maintaining production code while neglecting the maintenance of test code, which can lead to increased maintenance costs, such as those associated with regression testing.~\cite{pizzini2023sentinel}.
To illustrate the quality issues in test code, we analyze the Gson project~\cite{gson}, a popular Java library developed by Google, which has over 23.2k stars on GitHub and continues to be actively maintained, including its test code.
Figure~\ref{fig:moti} presents an intuitive example from the Gson's \textit{JsonObjectTest} class. 
According to the test smell types defined by \textit{tsDetect}~\cite{tsdetect}, this test has been identified with three distinct smells: 

\ding{182} \textbf{Eager Test}: This is a common and challenging type of smell to refactor. 
It occurs when a test method invokes several methods from the production code, making it harder to understand, maintain, and modify the test code effectively, as the test's purpose becomes less clear.
In this example, this test suffers from the \textit{Eager Test} smell because it is testing multiple behaviors—adding(\texttt{add}), removing(\texttt{remove}), and checking(\texttt{has} and \texttt{get}) properties—within a single test method. 
This violates the principle of \textit{Single Responsibility Principle (SRP)}~\cite{srp}. 
Each test method should focus on testing a specific behavior or function. 
Otherwise, it becomes harder to pinpoint the root cause of failures and reduce test clarity. 
To address this, each of these behaviors should be split into separate test methods to ensure that the test remains focused and maintainable.

\ding{183} \textbf{Duplicate Assert}: In this example, the repeated use of the same parameter \textit{propertyName} across multiple assertions leads to the identification of a \textit{Duplicate Assert} smell. 
Eliminating this smell can be achieved by either splitting the original test or transforming it into a \textit{Parameterized Test}, which removes the need for duplicate assertions by declaring parameterized values in an annotation and executing the test method multiple times.

\ding{184} \textbf{Assertion Roulette}: This is another common smell, indicating that multiple assertions in a test lack clear identification of the reasons for failure, and make it difficult to locate the failing assertion. 
To address this, each assertion should be supplemented with descriptive information that clarifies its purpose. 
This allows developers to quickly identify and fix issues when the test fails.

These test smells demonstrate the potential risks in test code that can compromise its quality and maintainability, highlighting the importance of systematically eliminating the existing smells.
Our tool, \appname, is designed to automatically refactor unit tests and eliminate test smells, thereby improving the quality of test code in existing software projects. 
Figure~\ref{fig:moti}(right) shows tests refactored using \appname, which have successfully removed the three identified test smells. 
Developers or testers can use \appname to refactor individual unit tests, single test files, or all tests in a project at once to eliminate smells and improve the overall quality of the software's code.

Another potential benefit of automated test refactoring lies in its impact on LLMs, many of which are trained on data from open-source communities.
By improving the quality of the test code that serves as training data, we can enhance the performance of LLMs in tasks such as test generation. 
For instance, during the data preprocessing phase, we can refactor test code that contains smells and use the cleaned, refactored code as training data. 
This approach may lead to more effective and reliable LLM-generated tests.
\section{approach}
\label{sec:approach}

\begin{figure}
\centering
\includegraphics[width=0.95\linewidth]{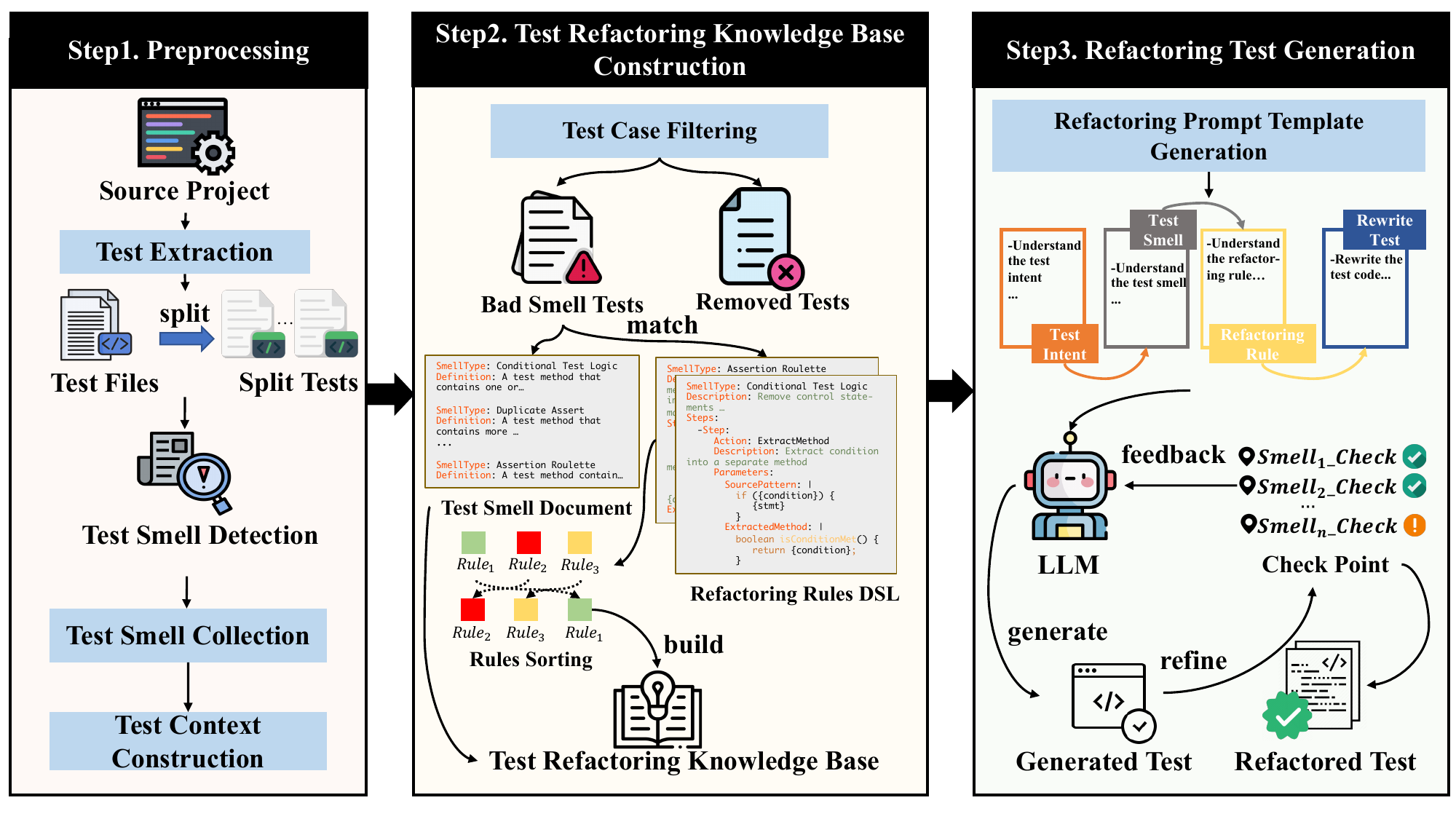}
\caption{Overview of our approach.}
\label{fig:approach}
\end{figure}

This section presents the details of our proposed approach \appname. 
As shown in Figure~\ref{fig:approach}, it can be divided into three main steps:

\textbf{Step} \ding{182} \textbf{Preprocessing.}
We begin by extracting the test code from the project and detecting all test smells present. 
For tests that need to be refactored due to the presence of smells, we also extract relevant code context (such as the tested methods and classes) during this step, treating this context as external knowledge for the LLM.

\textbf{Step} \ding{183} \textbf{Test Refactoring Knowledge Base Construction.} 
In this step, we design the types and definitions of test smells to serve as external knowledge that the LLM can reference. 
Additionally, we use a Domain-Specific Language (DSL) ~\cite{mernik2005and} to define corresponding refactoring rules for each type of test smell, constructing a comprehensive refactoring knowledge base.

\textbf{Step} \ding{184} \textbf{Refactored Test Generation.} 
This step focuses on generating the refactored test code. 
We design a specialized prompt template tailored for the test refactoring task. 
Moreover, we employ the Chain-of-Thought (CoT) ~\cite{cot} reasoning approach and propose a checkpoint mechanism to improve the LLM’s performance in refactoring test code.

\subsection{Preprocessing}
As shown in Figure~\ref{fig:approach}, this step involves three key processes: extracting the test code, identifying test smells, and gathering relevant contextual information about the tests.

\subsubsection{Test Extraction}
First, \appname automatically gathers all test files from the \texttt{test} directory (e.g., \texttt{src/test}) within a given Java project.
To avoid unnecessary detection of code smells and refactoring overhead, during the extraction process, we analyze whether the test files contain at least one method annotated with \texttt{@Test} and automatically filter out files that do not have the \texttt{@Test} annotation.
Besides, to analyze the focal methods in a fine-grained approach and meet the requirements of test smell detection, we collect the corresponding classes under test in this step.
Specifically, we strip the prefix and suffix \texttt{Test} from the test file names (e.g., \texttt{ParserTest}→\texttt{Parser}), and use this as an index to search for matching class files in the \texttt{src/main} directory. 
If a match is successful, the test file and the corresponding class under test are paired and collected together.

\begin{table}
\centering
\caption{Extracted test-related context information from the project.}
\begin{tabular}{l p{8cm}}
\toprule
\textbf{Items} & \textbf{Description} \\
\midrule

\textbf{Package Name} & The name of the package where the focal method is located, eg., \texttt{org.jsoup.parser} \\
\hline

\textbf{Focal Class} & The name of the class containing the focal method. 
\newline eg., \texttt{public class Parser} \\
\hline

\textbf{Focal Method Signature} & The signature of the focal method.
\newline eg., \texttt{public static Document parse(String html, String baseUri)} \\
\hline

\textbf{Focal Method Comment} & The code comment for the focal method.
\newline eg., \texttt{Parse HTML into a Document.} \\
\hline

\textbf{Other Invoke Methods} & The signatures of other methods invoked in the test.
\newline eg., \texttt{public static String unescapeEntities( String string, boolean inAttribute).} \\

\bottomrule
\end{tabular}
\label{tab:test_context}
\end{table}
  
\subsubsection{Test Smell Detection}
To detect test smells present in the project, we integrate the \textit{tsDetect}~\cite{peruma2020tsdetect}, an automated test smell detection tool for Java projects. 

\textbf{Refining Detection Granularity.} Our \appname performs refactoring at the level of individual test methods, refactoring one test at a time. 
However, the \textit{tsDetect} operates at a file-level granularity, meaning it can detect test smells in a test file as a whole but cannot pinpoint smells within specific test methods. 
To support the detection of smells at the level of individual test methods, we refine the test smell detection granularity.
Specifically, we employ a split-and-merge approach: before smell detection, we split each test file into multiple sub-files, with each sub-file containing a single test method and its necessary context.
This ensures that \textit{tsDetect} can analyze each test method independently. 
After the refactoring process is completed, these sub-files are merged back together into a complete test file based on the original split index.

\subsubsection{Test Context Collection}
Since refactoring operations are aimed at optimizing the code structure without altering its functional logic, it is crucial to ensure that the LLM has a sufficient understanding of the test code's functionality before refactoring.
Moreover, Yuan et al.~\cite{yuan2024evaluating} have demonstrated that providing additional test context can significantly enhance the performance of LLMs in test code generation tasks.
Inspired by these findings and applying them to the task of refactoring test code, we extract relevant contextual information from tests that exhibit smells and use it as external knowledge to help the LLM better comprehend the original test's intent, as shown in Table~\ref{tab:test_context}.

In this step, we extract the contextual information for each test that requires refactoring. 
During the subsequent refactoring steps, this information is integrated into the refactoring prompt template as a knowledge source to assist the LLM in accurately understanding the test's original intent.

\subsection{Test Refactoring Knowledge Base Construction}
In this step, we design the test smell types and corresponding refactoring rules as applicable external knowledge, aiming to enhance the LLM's understanding of test smells and effectively eliminate those present in unit tests.

\subsubsection{Test Smell Knowledge}
In the smell detection step, we integrate \textit{tsDetect}, which uses the 19 types of test smells defined by Peruma et al.~\cite{peruma2020tsdetect}. 
Although they define test smells and provide relevant code examples, the definition of test smells is not standardized across existing research, and the types and numbers of smells can vary between studies.

\begin{wrapfigure}{r}{0.52\textwidth}
\centering
\includegraphics[width=\linewidth]{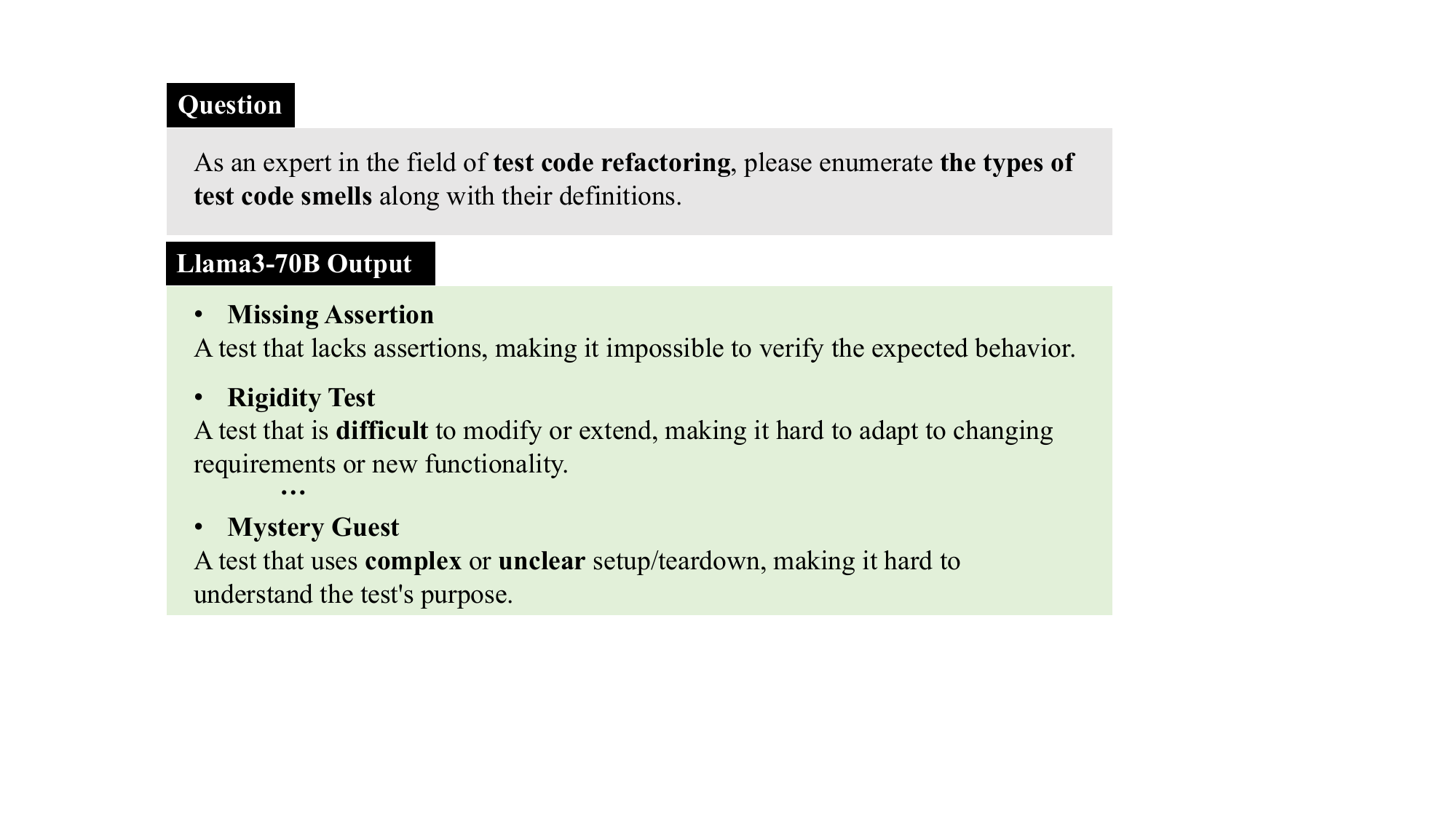}
\caption{An example of test smell explanation of the LLM.}
\label{fig:test_llm_exp}
\end{wrapfigure}

Additionally, LLMs are trained on data from various sources, such as blogs and GitHub repositories, which can lead to significant differences between LLMs in terms of training data.
This variability leads to inconsistencies in how LLMs interpret test smells, which in turn affects the effectiveness of the refactoring process. 
To illustrate this issue, we use \textit{LLaMA-70b}~\cite{llama3}. 
As shown in Figure~\ref{fig:test_llm_exp}, when asked about the meaning of specific test smells, the LLM’s responses deviate from our expectations. 
For example, \textit{tsDetect} does not include \textbf{Fragile Test} in its list of smells. 
Additionally, for the \textbf{Mystery Guest} smell, \textit{LLaMA-70b} provides a vague explanation, whereas \textit{tsDetect} defines it as the presence of unused variables in the \textit{setup/teardown} method.
There are also instances where the definitions are similar, but the naming conventions differ. 
For example, \textbf{Missing Assertion} refers to a test method that lacks a test oracle, but in \textit{tsDetect}, this is categorized as \textbf{Unknown Test}.

To mitigate the impact of these inconsistencies on the refactoring process, we provide the LLM with a clear and standardized set of test smell definitions before refactoring. 
Given that Peruma et al.~\cite{peruma2020tsdetect} provide comprehensive definitions, explanations, and code examples for 19 test smells in Java, we use these to build part of the external knowledge base for test smells, providing it to the LLM to eliminate inconsistencies in smell interpretation that could impact the refactoring process.
Due to space limitations, the detailed definitions of these smells are available in our replication package~\cite{utrefactor}.
For each smell type, we provide a structured definition, including a description of the test smell, its impact, and a pseudocode example to illustrate the concept.
During the refactoring process, these standardized smell definitions guide the LLM’s interpretation of the smells present in the test, ensuring consistent and accurate refactoring results.

\subsubsection{Test Refactoring Rule DSL}
To eliminate the smells present in tests, we design corresponding refactoring rules for each type of smell. 
We draw on previous manual approaches to eliminating test smells and empirical research on test smell refactoring~\cite{alfadel2023empirical}. 
Based on these insights and best testing practices, we formalize DSL rules for each type of test smell, which form the foundation of the entire refactoring process.
In designing these DSLs, we exclude simple smell types such as \textit{Default Test}, which refers to default test classes automatically created by \textit{Android Studio} when a project is initialized and do not require refactoring. 
Similarly, we handle \textit{Ignored Test} (tests marked as ignored) and \textit{Empty Test} (tests with empty method bodies) by simply removing the unit tests that exhibit these smells.
Finally, we define 13 method-level DSLs for Java test smells, aimed at eliminating smells in unit tests. 
Due to space limitations, the complete set of DSLs is available in our replication package~\cite{utrefactor}.
Since these DSLs determine the methods and steps for eliminating test smells, which is crucial for unit test refactoring, we evaluate the correctness of the DSL design by assessing whether the refactored tests successfully eliminate the smells without introducing new issues. 
This will be discussed in detail in subsequent sections.

\begin{wrapfigure}{r}{0.5\textwidth}
\centering
\includegraphics[width=\linewidth]{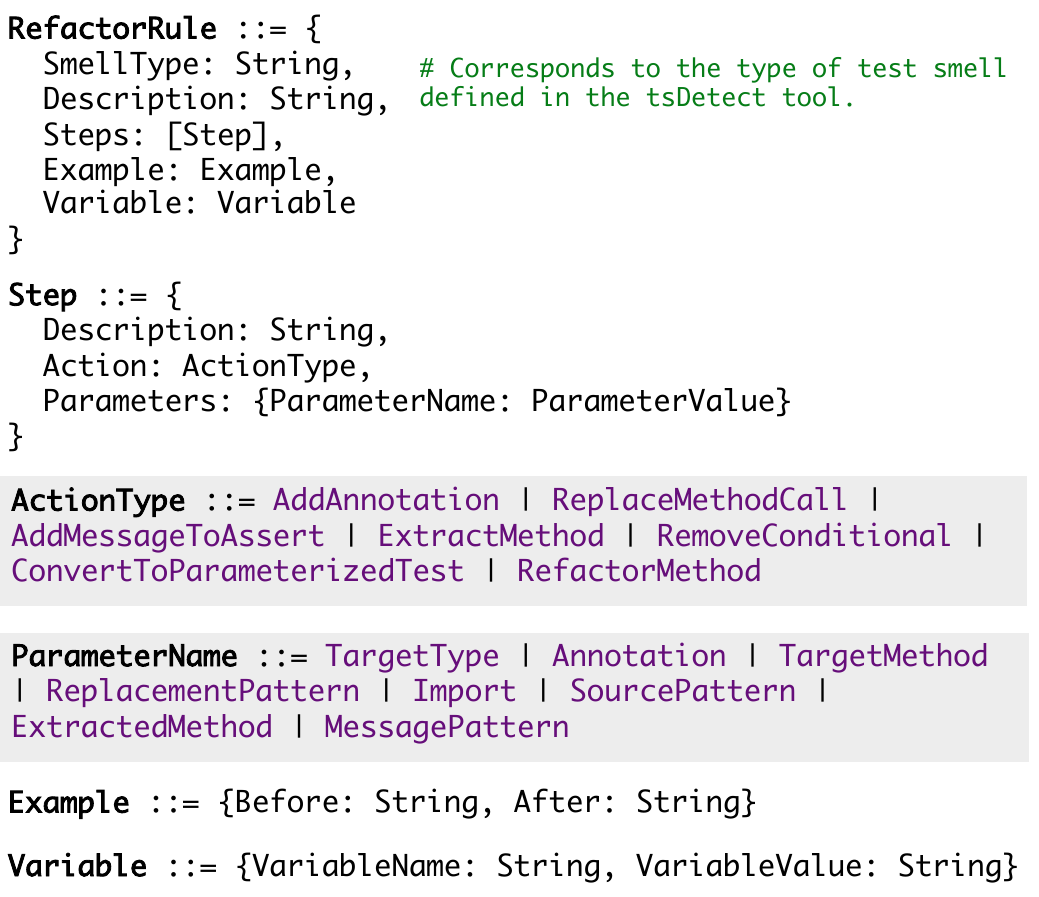}
\caption{A hierarchical definition of the DSL structure for test refactoring rules.}
\label{fig:dsl_structure}
\end{wrapfigure}

We leverage LLMs to refactor tests with identified smells according to the provided refactoring rules. 
A common approach is to describe these refactoring rules in natural language as prompts for the LLM. 
While this approach is straightforward and easy to understand, it presents challenges in practice: natural language can be ambiguous or imprecise, which may lead the LLM to generate code that does not meet expectations, especially when multiple smells are present in a single test. 
Without a clear strategy to guide the refactoring process, the LLM’s behavior can become unpredictable, resulting in inconsistent outcomes. 
Moreover, the lengthy natural language descriptions required for such complex cases can hinder the LLM's effectiveness in refactoring tests.

To address this issue, we propose using a more precise Domain-Specific Language (DSL) to express the test smell refactoring rules.
The advantage of using a DSL is that it allows for more precise and standardized descriptions of refactoring steps, thereby reducing the risk of misinterpretation.
Additionally, DSLs are typically structured, which facilitates the extension and maintenance of new refactoring rules. 
In practice, developers can easily add and update refactoring rules within the defined DSL structure. 
This structured approach ensures that refactorings are consistent, predictable, and less prone to errors, providing a clear framework for expanding the tool to accommodate new test smells and evolving best practices in refactoring.

\begin{figure}
\centering
\subfloat[The definition of DSL rules for eliminating AR (Assertion Roulette) smell]{
		\includegraphics[width=0.95\linewidth]{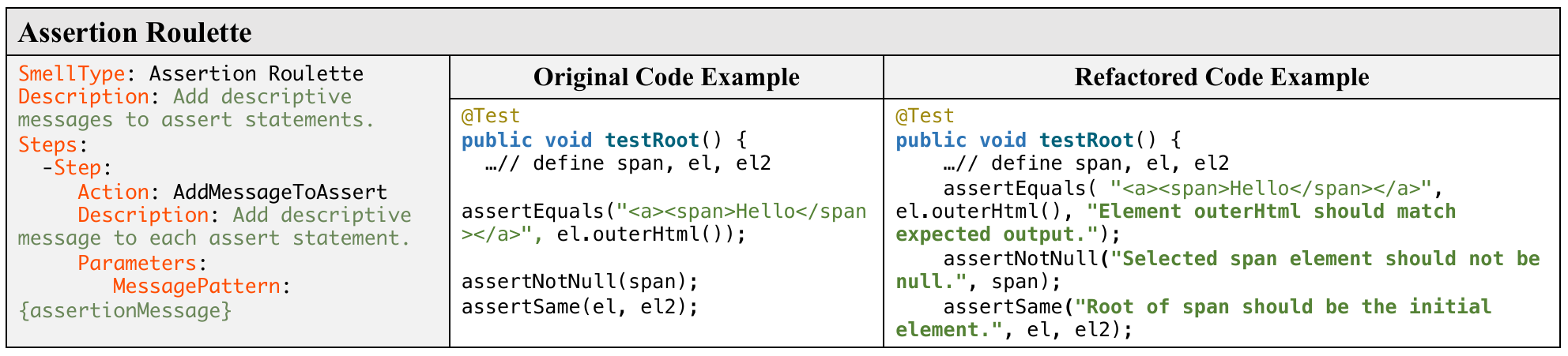}}\\
\subfloat[The definition of DSL rules for eliminating MNT (Magic Number Test) smell]{
		\includegraphics[width=0.95\linewidth]{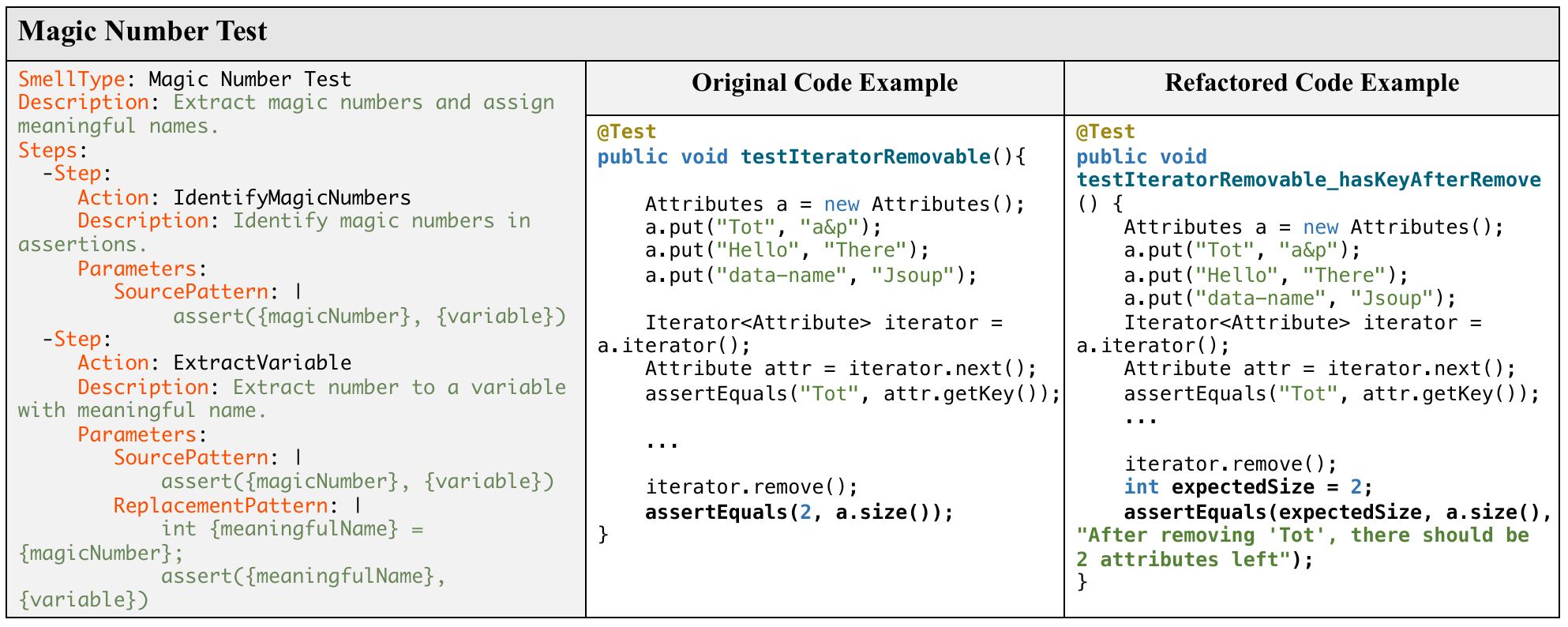}}\\
\subfloat[The definition of DSL rules for eliminating DA (Duplicate Assert) smell]{
		\includegraphics[width=0.95\linewidth]{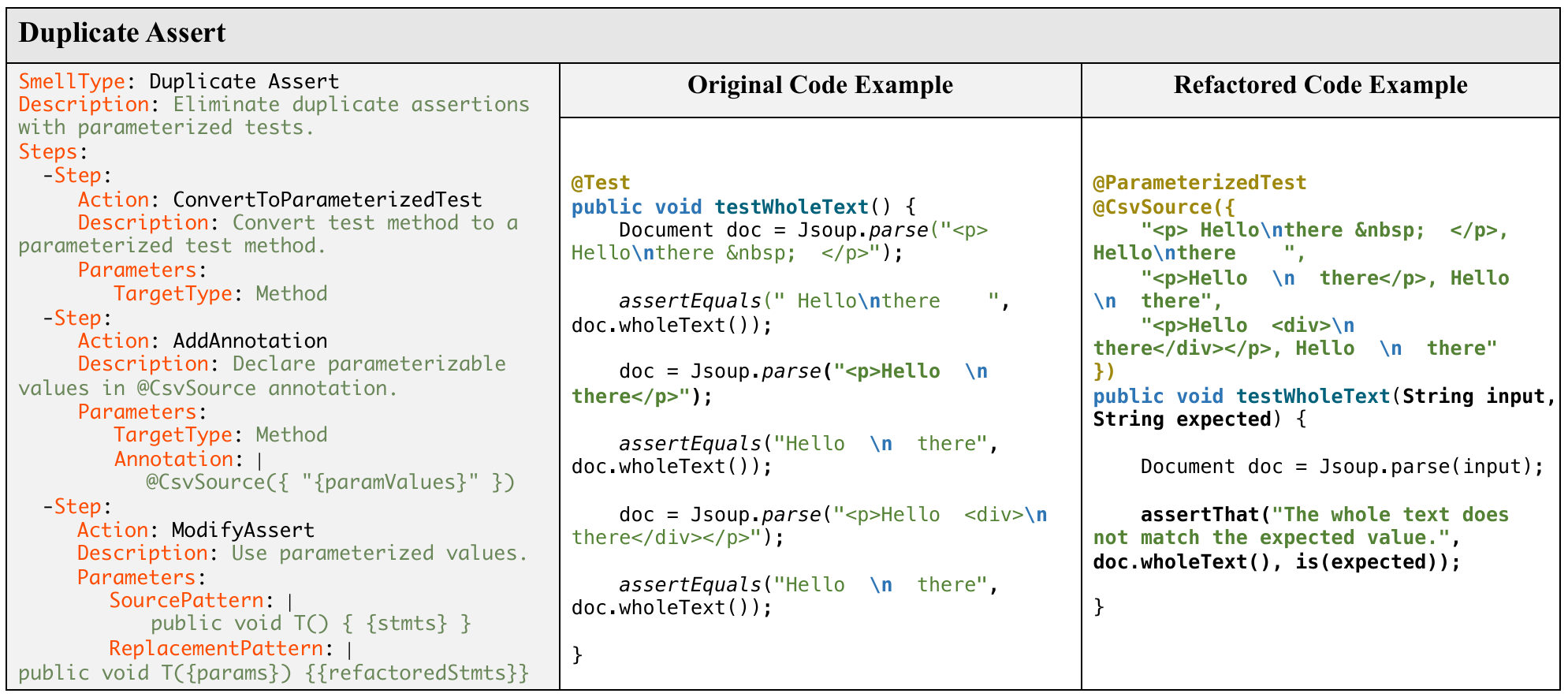}}
\caption{DSL rules for refactoring three types of test smells, accompanied by code examples illustrating the test code before and after refactoring.}
\label{fig:dsl_example}
\end{figure}


The DSL we design for test smell refactoring is illustrated in Figure~\ref{fig:dsl_structure}. 
Specifically, the \textit{SmellType} defines the specific category of test smell being addressed, ensuring that the appropriate refactoring strategy is applied. 
The \textit{Description} provides a concise explanation of how the refactoring will be executed for this smell type.
The \textit{Steps} component breaks down the refactoring process into a sequence of actions, enabling a fine-grained, step-by-step approach to modifying the test code. 
Additionally, the \textit{Example} provides before-and-after code snippets to demonstrate the practical application of the refactoring, aiding in both understanding and verification. 
Finally, \textit{Variables} parameterize the refactoring logic, offering flexibility and reusability across different test cases.
With this structure, the DSL systematically defines and executes complex code refactoring operations, which facilitates the automation of the test refactoring process.

Next, we present specific refactoring rules for test smells. 
Figure~\ref{fig:dsl_example} illustrates three concrete examples of these refactorings.
For example, when a test contains multiple assert statements without descriptive messages, it is flagged with the \textit{Assertion Roulette} smell. 
During refactoring, the test matches the DSL rule for the type \textit{Assertion Roulette}. 
This rule guides the refactoring process by specifying the action \textit{AddMessageToAssert}, which involves adding meaningful descriptive messages to each assert statement.
Another example is when a test is identified with the \textit{Duplicate Assert} smell. 
In this case, the test matches the corresponding DSL rule for \textit{Duplicate Assert}, which outlines a three-step refactoring process to eliminate the smell:  
\textbf{Step} \ding{182} Replace the \texttt{@Test} annotation with \texttt{@ParameterizedTest} to enable parameterized testing.
\textbf{Step} \ding{183} Add the \texttt{@CsvSource} annotation. This allows additional test cases to be added as parameters in \texttt{@CsvSource}, eliminating the need for multiple \texttt{assertEquals} statements.
\textbf{Step} \ding{184} Reduce duplicate code by writing the test logic only once. 
Parameterized testing makes the test cases more intuitive and clearly expresses the relationship between different inputs and expected outputs.

We put the details of DSLs on our replication package~\cite{utrefactor}.
These rules are then provided to the LLM as external knowledge during the refactoring process, guiding the LLM to refactor the test according to the predefined rules.

\subsubsection{Sorting Refactoring Rules}
\label{sec:sorting_rules}
Considering that a single test often exhibits multiple types of smells, it is crucial to determine the appropriate order of refactoring operations, as the sequence can significantly impact the effectiveness of the final test code refactoring. 
To address this, we categorize the test smell refactoring rules and assign them an execution priority. 
The LLM is then instructed to eliminate test smells according to their priority, from highest to lowest.

We have classified refactoring operations into three categories based on their characteristics: \textit{Removal}, \textit{Structural Optimization}, and \textit{Functional Optimization}. 
\ding{182} \textbf{Removal}: This category includes smells that require the test to be removed from the refactoring list rather than undergoing structural adjustments. 
Examples include \textit{EmptyTest} (test method with an empty body), \textit{Unknown Test} (test without assertions), and \textit{Default Test} (automatically generated default tests). 
Prioritizing removal operations helps to avoid unnecessary refactoring efforts for tests that do not contribute to code quality.
\ding{183} \textbf{Structural Optimization}: This category focuses on altering the existing structure of the test code. 
Examples include \textit{Eager Test}, where a test method calls multiple methods of the production object, and \textit{Duplicate Assertion}, where identical assertions are repeated within the same test method. 
\ding{184} \textbf{Functional Optimization}: This category targets improvements to assertion statements, such as \textit{Assertion Roulette} (where multiple assertions lack descriptive messages).

\textit{Structural Optimizations} take precedence over \textit{Functional Optimization} because a well-structured test provides a solid foundation for functional improvements.
For example, the \textit{Assertion Roulette} refactoring rule involves adding descriptive messages to assertions, and the \textit{Magic Number Test} refactoring rule extracts numeric literals into named variables. 
However, the \textit{Duplicate Assertion} refactoring rule eliminates duplicate assertions using parameterized tests, which could render the optimizations of \textit{Assertion Roulette} unnecessary if performed afterward. 
Therefore, it is crucial to refactor \textit{Duplicate Assertion} before applying other functional optimizations.

By following this prioritized refactoring approach, we ensure that the test code is improved in an orderly manner, starting with essential removals, followed by structural enhancements, and concluding with functional refinements. 

\subsection{Refactoring Test Generation}
In this final step, we generate the refactored test code using the specially designed prompt template. 
The template guides the LLM in applying the refactoring rules to eliminate the identified smells from the test code.

\subsubsection{Test Refactoring Prompt Design}
Typically, when a developer or tester refactors a test, they follow a process that involves: 
\ding{182} Understanding the intent of the test code, including what the focal method does and how the assertions verify its behavior.
\ding{183} Identifying the quality issues in the current test, specifically recognizing any smells present.
\ding{184} Most importantly, determining how to refactor the test to eliminate these smells.
\ding{185} Rewriting the test code accordingly.
Our approach is designed around these steps and is divided into four steps.
This four-step strategy is based on the Chain-of-Thought (CoT) paradigm, which enhances LLM reasoning capabilities by encouraging step-by-step thinking, leading to better outcomes in tasks such as extraction and reasoning~\cite{du2024vul}.

\begin{figure}
\centering
\includegraphics[width=0.95\linewidth]{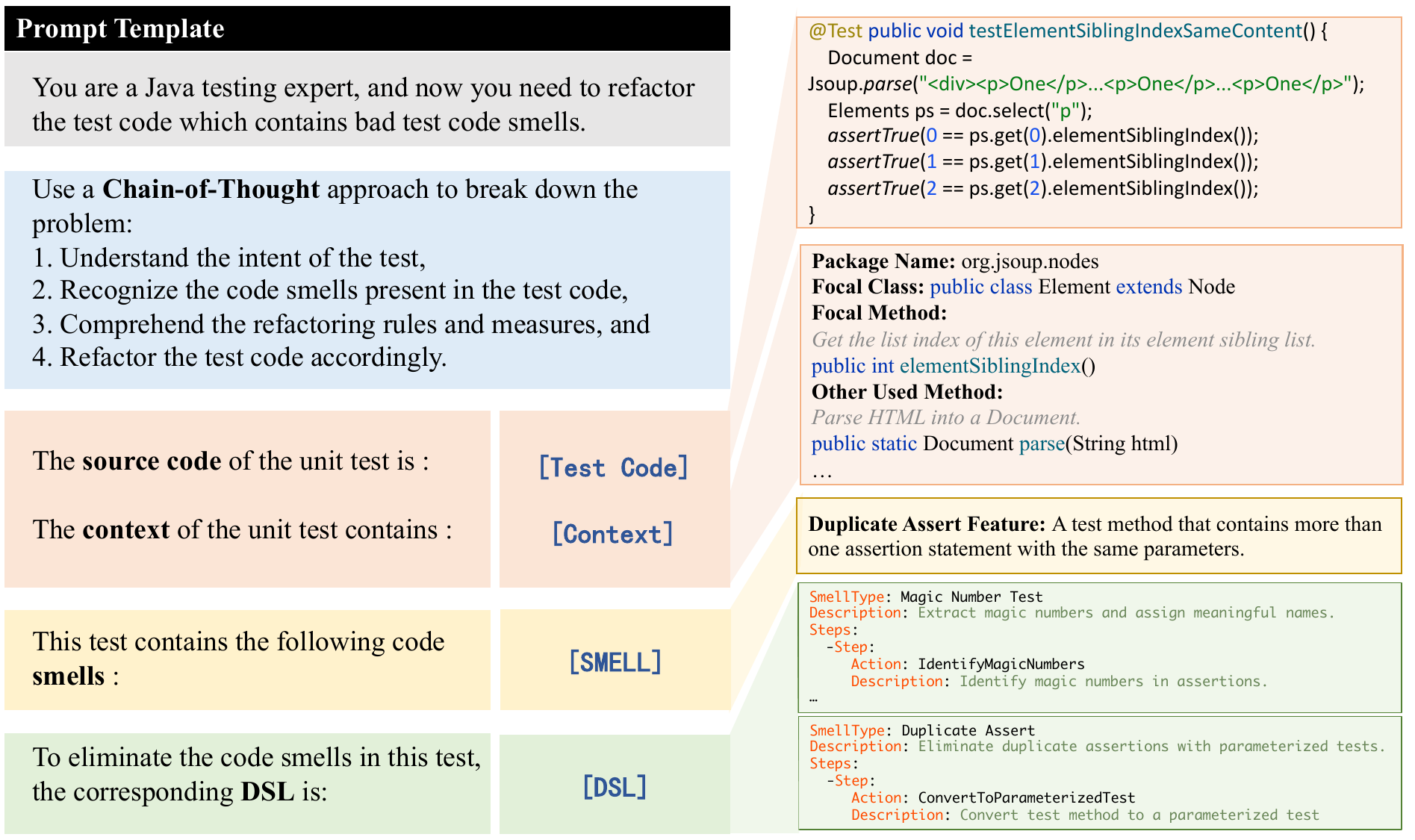}
\caption{The Prompt Template used for test smell elimination, along with corresponding examples of prompts.}
\label{fig:prompt_template}
\end{figure}

As illustrated in Figure~\ref{fig:prompt_template}, the complete prompt template based on CoT is designed to improve the LLM's ability to refactor tests. 
In guiding the LLM through the refactoring process, we simulate the actual steps a developer takes. 
First, to understand the test's intent, we provide the LLM with the test code along with relevant context, such as the tested class and method signatures. 
Next, we present the identified smells in the test and their corresponding definitions. 
Crucially, we supply the DSL-defined refactoring rules tailored to each smell type. 
Finally, with this external knowledge, the LLM proceeds to refactor the test.
This structured approach ensures that the LLM closely follows a logical and thorough process, much like a human developer or tester, resulting in more effective and accurate test refactoring.

\subsubsection{Checkpoint Mechanism}   
When a test contains multiple smells, even with sufficient information provided, the LLM may still miss eliminating some of the smells during refactoring. 
For example, in the \texttt{testSizeWhenHasInternal} method from the Jsoup project, the test has four identified smells: \textit{Assertion Roulette}, \textit{Eager Test}, \textit{Duplicate Assert}, and \textit{Magic Number Test}. 
Despite explicitly instructing the LLM to address all four types of smells in the prompt, in practice, the LLM often fails to fully refactor tests with multiple smells.

To address this issue and mitigate incomplete refactoring by the LLM, we design a checkpoint mechanism that allows the LLM to self-verify the completeness of its refactoring process. 
Specifically, for each smell in the test, we set up a checkpoint that asks whether the refactored test still exhibits the given issue. 
If the issue persists, the LLM continues refining the refactoring to eliminate the smell.
For tests with multiple smells, we construct a chain of smell checkpoints, as illustrated in Figure~\ref{fig:approach}. 
In the case of the test containing four smells, we collect the corresponding four checkpoints and integrate them into the prompt. 
The LLM uses this list to systematically check whether the smells have been fully removed. 
If any issues remain, the LLM will further adjust its refactoring until the test is completely free of the identified smells.
This checkpoint mechanism ensures that even in complex cases with multiple smells, the LLM can perform thorough and effective refactoring, ultimately leading to higher-quality test code.

\subsubsection{Refactoring Test Generation}  
It is essential for developers or testers to understand which tests have been refactored and the reasons behind these changes. 
To address this, we generate a detailed test refactoring report in this step. 
This report includes a list of refactored tests, specifying the file locations and test method names. 
It also outlines the detected smells for each test and the corresponding refactoring methods applied.
Additionally, to facilitate subsequent testing and compilation, we organize the refactored test files and methods according to the original project's test structure. 
\section{Evaluation}
\label{sec:evaluation}
Our experiments are designed to address the following research questions:

\begin{itemize}[leftmargin=*]
\item \textbf{RQ1: How is the quality of the test code after the refactoring of \appname?}
\item \textbf{RQ2: How effective is the refactoring of \appname?}
\item \textbf{RQ3: How effective is \appname in eliminating each category of Test Smell?}
\item \textbf{RQ4: What is the time efficiency of \appname in test refactoring?}

\end{itemize}

\subsection{Experimental Setup}

\noindent \textbf{Dataset.} 
We collect projects from GitHub, including those from well-known organizations (e.g., Apache~\cite{apache}) and popular projects (e.g., Jsoup~\cite{jsoup} with 10.9k stars).
To facilitate the test refactoring process, we set the following criteria for collection: first, to simplify test smell detection and make refactoring easier to validate, we collect projects written in Java and managed by Maven. 
Second, the source code of the project must compile successfully, and all unit tests are required to pass.
Finally, each project must exhibit at least 100 test smells as detected by the \textit{tsDetect}~\cite{tsdetect}.
Ultimately, we select six open-source projects for our test refactoring experiments.
As shown in Table~\ref{tab:datasets}, \appname's integrated tool \textit{tsDetect} identifies 879 unit tests with smells across these projects, with a combined total of 2,375 test smells detected.

\begin{table}
\centering
\caption{Test refactoring dataset (\textbf{\#TSmell} - number of tests with smells, \textbf{\#TS Num} - number of test smells).}
\begin{tabular}{lcccc}
\toprule
\textbf{Project} & \textbf{Tests} & \textbf{\#TSmell} & \textbf{\#TS Num} & \textbf{LOC} \\
\midrule
Commons-cli~\cite{cli} & 334 & 85 & 149 & 5,716 \\
Commons-compress~\cite{compress} & 1,585 & 37 & 263 & 29,249 \\
Commons-math~\cite{legacy} & 2,723 & 101 & 196 & 65,530 \\
Gson~\cite{gson} & 1,348 & 293 & 788 & 21,416 \\
Jfreechart~\cite{jfreechart} & 2,291 & 148 & 324 & 40,354 \\
Jsoup~\cite{jsoup} & 868 & 215 & 655 & 12,145 \\
\hline
\textbf{Total} & \textbf{9,149} & \textbf{879} & \textbf{2,375} & \textbf{174,510} \\
\bottomrule
\end{tabular}
\label{tab:datasets}
\end{table}

\noindent \textbf{Baselines.} 
For the baseline comparison of refactoring tools, we choose two tools for our unit test refactoring benchmarks.
First, \textit{TESTAXE}~\cite{de2022testaxe}, a tool implemented using the Rascal~\cite{rascal} meta-programming language, primarily designed to 
automatically upgrade projects from 
\textit{JUnit 4} to \textit{JUnit 5} to automatically detect and eliminate test smells in Maven-managed Java projects.
It is rule-based and leverages features in the \textit{JUnit 5} testing framework to remove test smells.
Second, considering that large language models (LLMs) inherently have the capability to refactor test code, we employ a default open-source LLM (\textit{Llama3-70B}) with a generic prompt as another comparison baseline.
As depicted in Figure~\ref{fig:prompt-example}, we use a standard prompt template, which provides the \textit{Llama3-70B} with the original test code and all detected test smell types. 
The LLM is then tasked with removing these smells and generating the refactored test code.

\begin{figure}[ht]
   \centering
   \begin{mdframed}[backgroundcolor=lightgreen!30, linecolor=gray, outerlinewidth=1pt]
    \textbf{Role:} \textit{You are a Java testing expert, and now you need to refactor the test code which contains bad test code smell.} \\
    \textbf{User:} Given the following \textcolor{blue}{Unit Test}, and the \textcolor{blue}{Test Smells} it contains, eliminate the smells present in this test without altering the original functionality and logic of the test, and provide the refactored test code. \\
    \textbf{Code and Smells:} \textbf{\textcolor{blue}{[Unit Test]}} \textbf{\textcolor{blue}{[Test Smells]}}
   \end{mdframed}
   \caption{Prompt Template for default LLM used to eliminate smell in tests.}
   \label{fig:prompt-example}
\end{figure}

\subsection{RQ1: How is the quality of the test code after the refactoring of \appname?}
Since the test refactoring process must eliminate test smells without altering the original test functionality, we first verify the quality of the refactored test code to ensure effective refactoring. 
This includes checking for code errors and ensuring that the functionality remains unchanged.
Specifically, we first assess the syntax correctness of the refactored test code. 
To achieve this, we calculate the compilation pass rate and test execution pass rate of all refactored tests, evaluating whether \appname introduces any errors during the structural changes that lead to compilation or runtime failures.

Table~\ref{tab:test_quality} presents the quality of the refactored test code across all open-source projects. 
As shown, \appname refactor 879 unit tests with identified test smells from six open-source projects.
The \textbf{\#CPR} column represents the compilation pass rate of the refactored tests. 
We manually compile all refactored tests and calculate that the average compilation pass rate is 95\%, with the highest rate being 97\% in the \textit{Jsoup} project.
\textbf{\#EPR} represents the execution pass rate of the refactored tests. 
Similarly, we manually execute all refactored tests and calculate an average \textbf{\#EPR} of 89\%, with the highest rate reaching 92\% in the \textit{Gson} and \textit{Jsoup} projects.
This indicates that while the refactoring process did introduce some errors, it remains largely effective, with the overall results demonstrating a solid level of reliability.

We analyze the tests where \appname introduces errors after refactoring. 
In the \textit{Gson} project, \texttt{testStrictComments} contains three assertions (\texttt{assertStrictError}) without descriptive information,  resulting in the detection of an AR (Assertion Roulette) smell.
During the refactoring process, \appname generates descriptive messages for each of the three assertions. 
However, a parameter mismatch error occurs during compilation.
Upon further analysis, we find that \texttt{assertStrictError} is a built-in assertion in \textit{Gson}, different from the standard \textit{JUnit} assertions. 
This assertion only accepts two parameters: an \texttt{exception} describing the expected behavior and an \texttt{expectedLocation} parameter. 
It does not support an additional message parameter. 
This conflicts with \appname's DSL rule for refactoring AR smells, which requires assertions to include a message parameter, leading to a compilation error and causing the smell to remain.

We calculate the coverage of tests that still pass after refactoring by using the \textit{Jacoco}~\cite{jacoco} test coverage tool in each project.
As shown in Table~\ref{tab:test_quality}, we observe that \appname does not alter the original line coverage or branch coverage of the tests after refactoring.
This indicates that \appname only restructures the code, such as refactoring assertions or breaking down test methods, without changing the original functionality. 
It avoids introducing new calls to the project's production APIs or omitting existing conditions or function calls within the tests.
This outcome aligns with our expectations, as the goal of \appname is to eliminate test smells and improve test code quality, rather than modifying the underlying functionality.

\begin{table}
\centering
\caption{Quality of refactored test code generated by \appname (\textbf{\#RTests} - number of tests after refactoring, \textbf{\#CPR} - test compilation pass rate, \textbf{\#EPR} - test execution pass rate).}
\begin{tabular}{l|c|c|c|c|c}
\hline
\multirow{2}{*}{\textbf{Project}} & \multirow{2}{*}{\textbf{\#RTests}} & \multirow{2}{*}{\textbf{\#CPR}} & \multirow{2}{*}{\textbf{\#EPR}} & \textbf{Line Coverage} & \textbf{Branch Coverage} \\
\cline{5-6}

& & & & \textbf{cov.}(\textbf{unchgd.}) & \textbf{cov.}(\textbf{unchgd.})  \\
\hline

Commons-cli & 126 & 93\% & \textbf{89\%} & 98\% (\checkmark) & 95\% (\checkmark) \\
\hline

Commons-compress & 65 & 92\% & \textbf{83\%} & 84\% (\checkmark) & 75\% (\checkmark)\\
\hline

Commons-math & 144 & 95\% & \textbf{88\%} & 92\% (\checkmark) & 85\% (\checkmark) \\
\hline

Gson & 469 & 94\% & \textbf{92\%} & 87\% (\checkmark) & 85\% (\checkmark) \\
\hline

Jfreechart & 314 & 95\% & \textbf{81\%} & 54\% (\checkmark) & 45\% (\checkmark) \\
\hline

Jsoup & 404 & 97\% & \textbf{92\%} & 86\% (\checkmark) & 80\% (\checkmark) \\
\hline

\textbf{Total} & \textbf{1,522} & \textbf{95\%} & \textbf{89\%} & \textbf{77\%} (\checkmark) &\textbf{67\%} (\checkmark) \\
\hline
\end{tabular}
\label{tab:test_quality}
\end{table}

\textbf{Consistency of Refactoring}. Given the challenge of determining functional and behavioral consistency of test code before and after refactoring, we employ a more advanced model (GPT-4o) to evaluate the functional consistency of test code pre- and post-refactoring, supplemented by a manual secondary review of the evaluation results. 
Specifically, we provide GPT-4o with all 879 pairs of test code that can be executed successfully before and after refactoring to assess whether the changes involve only structural adjustments while maintaining functional consistency. 

\begin{wraptable}{l}{0.45\textwidth}
\centering
\caption{Classification of test inconsistencies before and after refactoring.}
\begin{tabular}{l|c}
\hline
\textbf{Inconsistent Type} & \textbf{Num} \\
\cline{1-2}
Exception Handling & 18 \\
\hline
Redundant Assertion & 9 \\
\hline
Redundant Variable Declaration & 5 \\
\hline
\end{tabular}
\label{tab:correct}
\end{wraptable}

Table~\ref{tab:correct} indicates that 32 cases (approximately 4\%) are flagged as \textbf{Inconsistent}, which can be categorized into three types: Firstly, regarding differences in exception handling (18 cases), although GPT-4o identifies a behavioral change between the original test code's throws exception handling and the DSL-refactored JUnit \texttt{assertDoesNotThrow} method, this refactoring actually aligns with the design philosophy of the JUnit 5 framework, providing a more standardized exception verification mechanism.

Secondly, in terms of redundant assertion handling (9 cases), the current DSL strategy of directly deleting redundant assertions is marked by GPT-4o as a potential behavioral change, prompting us to plan an optimization of the refactoring rules to comment out redundant assertions and add developer prompts. 
Lastly, concerning redundant variable declarations (5 cases), variables defined during the setup phase are redundantly declared in subtests when splitting the original test into multiple subtests. 
Although there is still room for improvement in \appname's handling of redundant assertions and duplicate variables, it performs excellently overall in maintaining test functional consistency.

\subsection{RQ2: How effective is the refactoring of \appname?}
We are next interested in evaluating the extent to which \appname eliminates bad smells in refactored test code that still passes execution, thus validating the effectiveness of \appname in test refactoring. 
Additionally, we compare and analyze \appname's effectiveness against two baseline tools.
To assess refactoring effectiveness, we focus on two key metrics in this research question (RQ): 
the number of passing tests after refactoring and the number of remaining smells in those tests. 
We also measure the change in the total number of smells before and after refactoring.
A higher reduction rate indicates greater effectiveness of our automated refactoring approach in minimizing test smells across the project.

Table~\ref{tab:experiment_results} shows the effectiveness of \appname and the two baseline tools in refactoring tests across six Java open-source projects.
Out of the 2,375 test smells present across the six projects, \appname reduced the number to 265 after refactoring, achieving an average reduction rate of 89\%. 
The \textit{Commons-compress} project shows the highest reduction rate, reaching 94\%.
This demonstrates that \appname is highly effective in reducing test smells.
In particular, for the most prevalent smell, AR (Assertion Roulette), \appname had the most significant impact, completely eliminating this smell by successfully adding descriptive messages to all assertions that lacked them.
The effectiveness of \appname in addressing different types of test smells will be discussed further in RQ3.

\textbf{Comparison with \textit{TESTAXE}.} As shown in Table~\ref{tab:experiment_results}, \textit{TESTAXE} achieves a 19\% smell reduction rate in the \textit{Commons-math} project, while in the \textit{Commons-compress} and \textit{Jsoup} projects, the reduction rate is less than 1\%. 
In the \textit{Commons-cli}, \textit{Gson}, and \textit{Jfreechart} projects, \textit{TESTAXE} fails to eliminate any test smells.
\textit{TESTAXE} primarily aims to upgrade projects from JUnit 4 to JUnit 5. 
During this process, it uses JUnit 5 features to refactor existing tests and eliminate test smells.
\textit{TESTAXE} relies on a set of built-in syntax matching and replacement rules for test refactoring. 
However, it includes only five rules related to smell elimination, limiting its effectiveness. 
For example, it cannot handle AR (Assertion Roulette), one of the most common test smells. 
Additionally, \textit{TESTAXE}'s replacement rules require strict matching with the original test code, meaning that any code not covered by these rules remains unchanged.
The biggest limitation of their approach is its poor adaptability. For example, in handling ECT (Exception Catching Throwing) smells, \textit{TESTAXE} uses the \texttt{ExpectedExceptionTransformation} rule, which only matches patterns like \texttt{throws} in function signatures. 
However, test cases often involve exceptions handled in \texttt{try-catch} blocks that require refactoring, and since \textit{TESTAXE}'s rules do not cover these patterns, it cannot eliminate such exception smells.
These limitations significantly reduce \textit{TESTAXE}'s time efficiency in eliminating and refactoring test smells.

\begin{table}
\centering
\caption{Comparison results of test smell elimination effectiveness(\textbf{\#TS Num} - number of Test Smells).}
\begin{tabular}{l|c|c|c|c|c|c|c}
\hline
\multirow{2}{*}{\textbf{Project}} & \multirow{2}{*}{\textbf{\#TS Num}} & \multicolumn{2}{c|}{\textbf{\appname}} & \multicolumn{2}{c|}{\textbf{Llama3-70B}} & \multicolumn{2}{c}{\textbf{TESTAXE}} \\
\cline{3-8}
& & \textbf{aft.} & \textbf{rate} & \textbf{aft.} & \textbf{rate} & \textbf{aft.} & \textbf{rate} \\
\hline
Commons-cli & 149 & 21 & $\downarrow$ \textbf{86\%} & 94 & $\downarrow$ 37\% & - & - \\
\hline
Commons-compress  & 263 & 16 & $\downarrow$ \textbf{94\%} & 55 & $\downarrow$ 79\% & 261 & $\downarrow$ <1\% \\
\hline
Commons-math  & 196 & 28 & $\downarrow$ \textbf{86\%} & 107 & $\downarrow$ 45\% & 158 & $\downarrow$ 19\% \\
\hline
Gson  & 788 & 87 & $\downarrow$ \textbf{89\%}  & 375 & $\downarrow$ 52\% & - & - \\
\hline
Jfreechart & 324 & 44 & $\downarrow$ \textbf{86\%} & 197 & $\downarrow$ 39\% & - & - \\
\hline
Jsoup & 655 & 69 & $\downarrow$ \textbf{91\%} & 252 & $\downarrow$ 62\% & 635 & $\downarrow$ <1\% \\
\hline
\textbf{Total} & \textbf{2,375} & \textbf{69} & $\downarrow$ \textbf{89\%} & \textbf{1,080} & $\downarrow$ \textbf{55\%} & \textbf{2,315} & $\downarrow$ \textbf{<1\%} \\
\hline
\end{tabular}
\label{tab:experiment_results}
\end{table}

\textbf{Comparison with \textit{Llama3-70B}.} It is evident that when only providing the test code and corresponding smell types to \textit{Llama3-70B} for refactoring, it reduces a total of 1,080 smells across the six projects, with an average reduction rate of 55\%. 
This is lower than that achieved by \appname.
This shows that while the LLM has the ability to refactor test smells, its effectiveness is limited.
After manual inspection of the LLM-refactored tests, we find that it performs best in handling AR (Assertion Roulette) smells. 
In tests with AR smells, the LLM successfully adds descriptive messages to assertions that lack them, which is attributed to its inherent code understanding capabilities. 
However, it does not eliminate all AR smells. 
When multiple smells are present in a test, using the default prompt (as shown in Figure~\ref{fig:prompt_template}) often leads to incomplete smell elimination, leaving AR smells unresolved.
This issue is mitigated in \appname by using a checkpoint mechanism, which ensures that each smell is checked and eliminated after refactoring.
For other types of smells, the default LLM's ability is also limited. 
For example, in handling exception-related smells, without the structured refactoring DSL rules designed in \appname, the  \textit{Llama3-70B} often tends to simply replace \texttt{throws} statements with \texttt{try-catch} blocks, which does not effectively remove the smell.
Overall, the default LLM exhibits more randomness in test refactoring. 
This characteristic becomes more prominent when multiple smells are present in a single test, resulting in unpredictable refactoring outcomes. 
In contrast, \appname, with its integration of external knowledge and structured refactoring rules, consistently delivers more effective and stable smell elimination.

\textbf{Effectiveness of Assertion Roulette Elimination}. Adding any string message to assertions can eliminate the Assertion Roulette (AR) smell. 
However, the generated messages should clearly convey the intent of the assertion. 
To evaluate the quality of these messages, we conduct an experiment using GPT-4o, which is tasked with evaluating messages based on two criteria: \ding{182} \textbf{Is the message clear and explicit?} \ding{183} \textbf{Does the message align with the intent of the assertion? }
In this experiment, we collect all 1,230 generated assertion messages. 
The results show that only 43 messages (3.5\%) are flagged as \textbf{Unclear}, and all of these messages fail to meet criterion \ding{182} (clarity of the message). 
After manual review, we find that these cases share a common feature: they lack sufficient context. 
For example, numerical variables (such as Magic Numbers) are extracted as constants with meaningful names. 
\texttt{assertEquals(expectedDatasetIndex, li.getDatasetIndex(), "The dataset index of the legend item should be 1")}, GPT-4o marks the message as unclear because it misinterprets \texttt{expectedDatasetIndex} as a variable rather than a constant. 
However, the generated message is reasonable, as the intent behind it is clear.
Therefore, we conclude that LLMs can generate clear and descriptive assertion messages that align with the intent of the assertion.

\textbf{Ablation of the DSL}. We conduct an ablation study by comparing the effectiveness of \appname with a version excluding the DSL, while keeping all other aspects the same (e.g., detailed descriptions of test smells, test context, and the smell checkpoint mechanism).
The results show that without the DSL the reduction rate of test smells is 67\%, which is an improvement over using LLM alone (55\%) but significantly lower than the 89\% achieved with the DSL.

Additionally, the natural language descriptions of test smells allow the LLM to partially address them effectively. For example, in 509 cases of Exception Catching Throwing, 175 cases (34\%) successfully eliminate the smell using \texttt{assertDoesNotThrow} or \texttt{assertThrow}. 
However, the remaining cases show randomness, such as removing try-catch blocks or replacing throws with try-catch. 
In contrast, the DSL clearly defines rules for eliminating such smells, covering different scenarios (e.g., adding exception descriptions), leading to more consistent results.

Finally, we analyze all refactored tests that fail at runtime after successful compilation. 
Most failures occur when handling Duplicate Assertions, as our DSL refactors the original tests into JUnit 5 parameterized tests. 
The LLM occasionally introduces errors in the parameterized data provided via \texttt{@CsvSource}, causing runtime failures. 
We find that the LLM struggles with converting tests into the parameterized format. 

\begin{figure}
\centering
\includegraphics[width=0.95\linewidth]{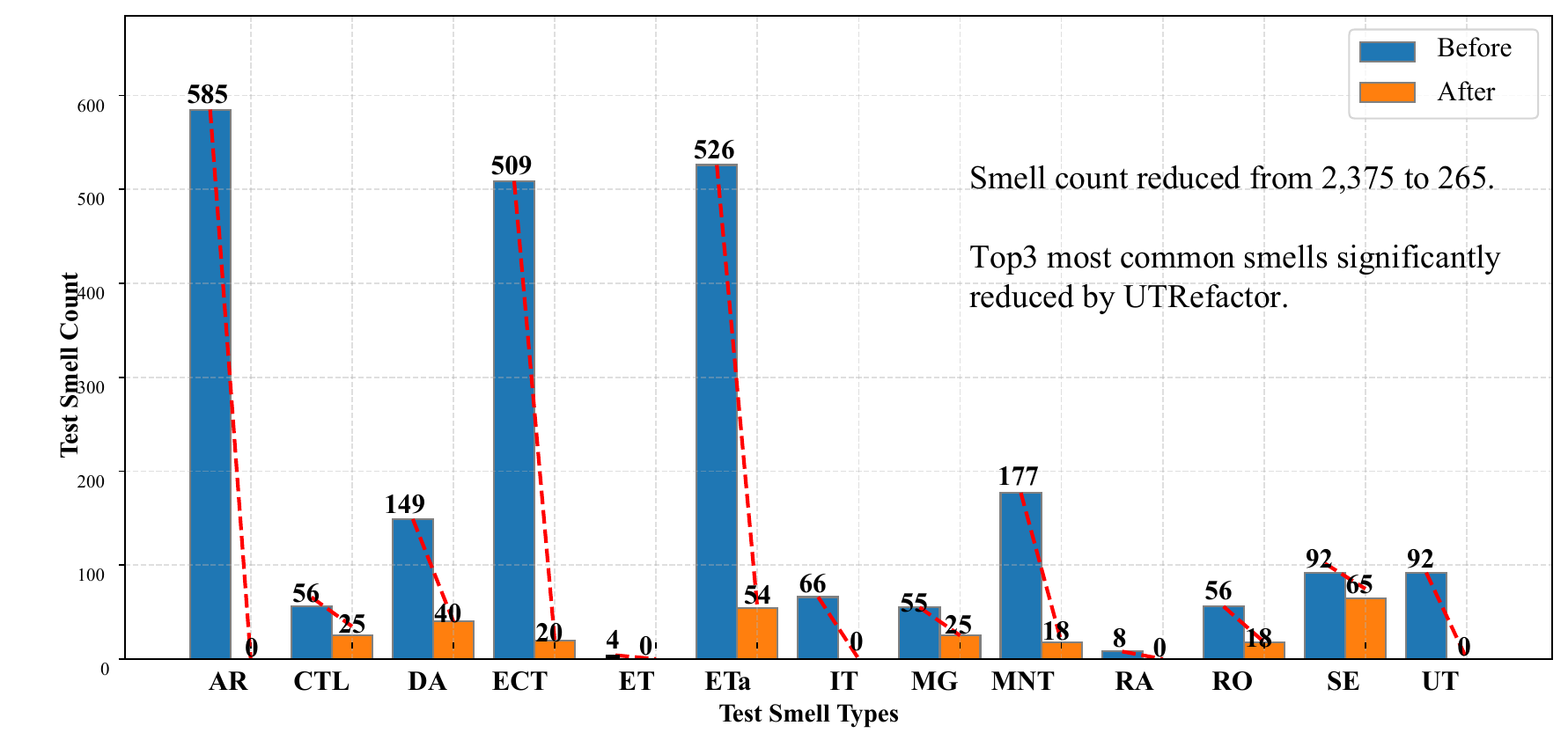}
\caption{Changes in the number of each type of test smell before and after test refactoring in \appname(\textbf{MG}=Mystery Guest, \textbf{RO}=Resource Optimism, \textbf{MNT}=Magic Number Test, \textbf{SE}=Sensitive Equality, \textbf{ETa}=Eager Test, \textbf{DA}=Duplicate Assert, \textbf{CTL}=Conditional Test Logic, \textbf{ECT}=Exception Catching Throwing, \textbf{AR}=Assertion Roulette, \textbf{RA}=Redundant Assertion, \textbf{IT}=Ignored Test, \textbf{UT}=Unknown Test, \textbf{ET}=Empty Test).}
\label{fig:smell_num_change}
\end{figure}

\subsection{RQ3: How effective is \appname in eliminating each category of Test Smell?}
In this research question, we focus on \appname's ability to eliminate different types of test smells. 
Refactoring based on the type of code smells can lead to the same test being refactored multiple times, as many smells often coexist within a single test. 
Additionally, the order in which refactoring is performed can influence the final outcome (sec~\ref{sec:sorting_rules}). 
To address these challenges, our refactoring process targets individual tests and simultaneously eliminates all smells.
We first detect and collect all types of test smells present in the six open-source projects. 
We then analyze how effectively \appname removes these smells.

As noted in RQ2, \appname successfully reduces the total number of test smells from 2,375 to 265.
Figure~\ref{fig:smell_num_change} provides detailed data on \appname's ability to eliminate each type of test smell. 
From the figure, we observe that \appname completely eliminates the most prevalent type, i.e., AR (Assertion Roulette) smell. 
This success is due to \appname's integration of LLM code understanding capabilities, enabling it to infer appropriate descriptive messages based on the specific assert statements and surrounding code context. 
These messages help developers quickly identify the cause of errors when tests fail. 
Additionally, \appname shows significant refactoring results for the two other high-frequency smells, ECT (Exception Catching Throwing) and ETa (Eager Test), reducing their numbers from 526 and 509 to 54 and 20, respectively.

We manually review the refactored tests generated by \appname and analyze its effectiveness in eliminating each type of smell. 
Taking ETa (Eager Test) as an example, this smell occurs when a test includes assertions for multiple production functions, making the test’s purpose unclear and violating the best practice of the single responsibility principle. 
The refactoring approach for this smell is to split the test based on the production functions it verifies.
\appname's advantage in eliminating ETa is demonstrated in two ways. 
First, its ETa DSL clearly defines how to identify API assertions, split the test, and generate new test methods.
Second, with the help of LLM, \appname can understand and reason based on context, effectively handling complex code structures while avoiding the randomness typically associated with LLM-generated tests.

\begin{wrapfigure}{r}{0.5\textwidth}
\centering
\includegraphics[width=\linewidth]{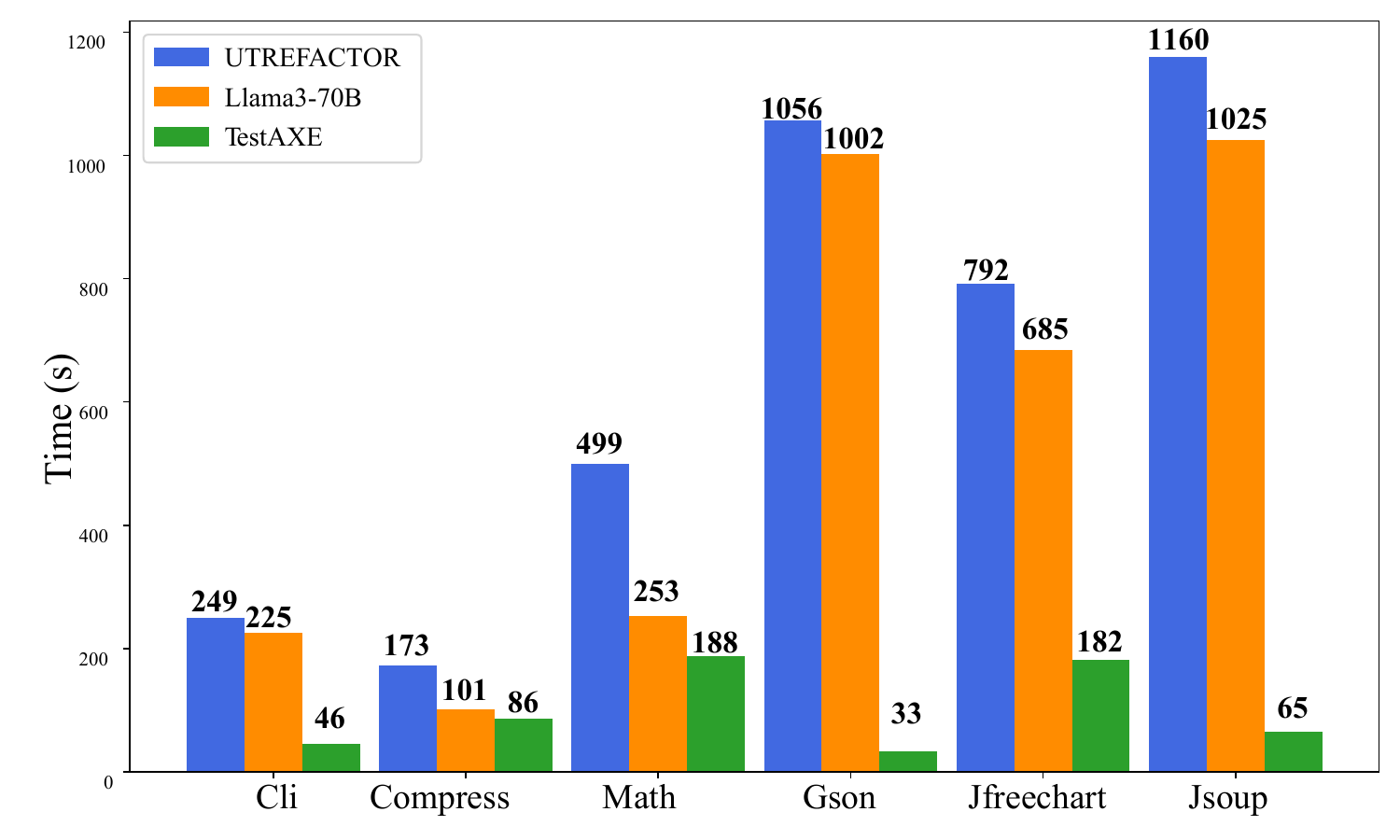}
\caption{Comparison of time costs between \appname and two baseline tools across each project.}
\label{fig:time_compare}
\end{wrapfigure}

For the cases where smell elimination failed, we conducted a manual review and analysis. 
From the Figure~\ref{fig:smell_num_change}, \appname's ability to handle certain less frequent smells is somewhat limited. 
A notable example is SE (Sensitive Equality), which occurs when the default \textit{toString} method is used in assertions. 
The refactoring approach for this smell requires either the project to have overridden the \textit{toString} method or to provide an equivalent method for object comparison.
In other words, if the project neither overrides \textit{toString} nor offers a functionally equivalent method, \appname is unable to remove this type of smell. 
Overall, across the six open-source projects, \appname shows less effectiveness in addressing this smell type, primarily due to the lack of overridden \textit{toString} methods or alternative methods in the project.

\subsection{RQ4: What is the time efficiency of \appname in test refactoring?}
In this research question, we focus on the time efficiency of \appname during the test refactoring process. 
We first collect the time data for each step of \appname and then compare the total time with the two baseline tools.

\begin{wraptable}{l}{0.45\textwidth}
\centering
\caption{Time consuming(s) for each step of refactoring tests in \appname.}
\resizebox{0.45\textwidth}{!}{\begin{tabular}{lcccc}
\toprule
\textbf{Project} & \textbf{Step1} & \textbf{Step2} & \textbf{Step3}  \\
\midrule
Cli & 24 & 4 & 221 \\
Compress & 69 & 3 & 101 \\
Math & 258 & 6 & 235\\
Gson & 64 & 4 & 988  \\
Jfreechart & 90 & 5 & 697  \\
Jsoup & 31 & 2 & 1,127  \\
\hline
\textbf{Total} & \textbf{536 (12\%)} & \textbf{24 (<1\%)} & \textbf{3,369 (87\%)} \\
\bottomrule
\end{tabular}}
\label{tab:step_time}
\end{wraptable}

Table~\ref{tab:step_time} illustrates the time spent by \appname at each refactoring step. 
Across all six open-source projects, the total time from detecting test smells to refactoring the 879 tests with identified smells is 4,379 seconds, averaging 3.8 seconds per test. 
This highlights \appname's efficiency in terms of time consumption.
Moreover, to accommodate various real-world refactoring scenarios, such as automatically refactoring test smells immediately after developers write a test or refactoring the entire test code of an existing project, \appname supports refactoring at different levels: a single test, a single test file, or all test files within a project.

\section{Discussion}
\label{sec:discussion}
\subsection{Strengths of \appname}
In \appname, we explore how LLMs perform automatic test refactoring tasks, guided by rules described in DSLs. 
This approach simulates the process that developers or testers follow during refactoring, from understanding the test's intent and identifying test smells to following step-by-step instructions outlined in the DSL. 
Compared to directly using an LLM for test refactoring, \appname significantly reduces the risk of random errors, which can make the refactoring process unpredictable.

To further illustrate this issue, we randomly select 100 cases where the LLM fails to eliminate smells and conduct an analysis. 
Among them, 65 cases fail due to the LLM's arbitrary refactoring behavior, such as removing throws declarations or try-catch blocks without resolving the Exception Catching Throwing smell.
22 cases fail due to hallucinations during the elimination of the Eager Test smell, where the LLM introduces non-existent functions or variables into split tests.
23 cases fail because the LLM generates helper methods with empty bodies and comments instructing users to complete them.


\subsection{Integration of \appname with LLMs}
Closed-source LLMs often come with high token usage costs, and their large parameter sizes lead to significant deployment expenses. 
To address this, \appname uses the open-source and free \textit{Llama3-70B} model.
However, considering the tool's extensibility, \appname also supports replacing \textit{Llama3-70B} with other LLM models. 
\textit{ChatGPT}~\cite{chatgpt} is a widely recognized model and the current stable version \textit{GPT-4o}~\cite{gpt4}.
We design an experiment to demonstrate that \appname effectively eliminates test smells using alternative LLM models.
As shown in Table~\ref{tab:gpt4}, we compare two setups: one using the default \textit{GPT-4o} with the general prompt (as illustrated in Figure~\ref{fig:prompt-example}) and another using \appname with \textit{Llama3-70B} replaced by \textit{GPT-4o}.
We randomly select 100 tests with smells from six open-source datasets, containing a total of 251 test smells. 
After refactoring the tests, Table~\ref{tab:gpt4} shows that using \textit{GPT-4o} with the default prompt achieves a smell reduction rate of 55\%, whereas \appname achieves a reduction rate of 91\%.

\begin{wraptable}{l}{0.45\textwidth}
\centering
\caption{Effectiveness of \appname integrated with GPT-4o in eliminating test smells.}
\resizebox{0.45\textwidth}{!}{\begin{tabular}{l|c|c}
\hline
\textbf{Tool} & \textbf{\#TS Num} & \textbf{aft.(rate)}  \\
\cline{1-3}
Default GPT-4o & 251 & 108 ($\downarrow$57\%)  \\
\hline
\appname(GPT-4o) & 251 & 23 ($\downarrow$91\%)  \\
\hline
\end{tabular}}
\label{tab:gpt4}
\end{wraptable}

We manually review the test refactoring results using \textit{GPT-4o}. 
Without guidance from external knowledge on test smells, the default \textit{GPT-4o} encounters issues in eliminating smells. 
For example, in the \texttt{testFindRangeBounds} of the \textit{jfreechart}, where four types of smells are present.
\textit{GPT-4o} removes only three smells but fails to eliminate the most common smell, \textit{Assertion Roulette}. 
In other cases, we observe that \textit{GPT-4o} attempts to address \textit{Assertion Roulette} by adding comments to assertion statements instead of explicitly adding a message, which does not properly resolve the smell.

In contrast, \appname leverages its DSL to clearly instruct the addition of a message to assertions and ensure that all smells are addressed during refactoring, avoiding the issues caused by the default LLM approach.
In another example, the \texttt{testDrawWithNullInfo} contains a complex \texttt{try/catch} structure detected as \textit{Exception Catching Throwing}. 
\textit{GPT-4o} simply removes this block and adds an explicit \texttt{throw} in the method signature, but this fails to eliminate the smell. 
By contrast, guided by \appname's DSL, the tool uses \texttt{assertNotThrow} to replace the \texttt{try/catch} block and successfully resolves the smell.
This demonstrates that \appname effectively eliminates smells even when using other LLMs like \textit{GPT-4o}.

\subsection{Threats to Validity}
\textbf{Threats to internal validity.} To minimize hallucinations and reduce errors during the test smell elimination process, we provide LLMs with sufficient context related to test refactoring and design a refactoring DSL to guide the LLM step by step in removing test smells. 
However, the DSL in \appname may not cover every possible scenario, which can impact the overall effectiveness of the test refactoring.
Our refactoring DSL is designed as a flexible, external configuration file, making it easy to extend. 
For example, new smell types can be addressed by simply adding or updating DSL files, modifying refactoring steps to accommodate a broader range of scenarios.

\textbf{Threats to external validity.} 
We integrate the \textit{tsDetect} tool into our process for test smell detection.
This tool uses a built-in Java parser to analyze test code and identify smells, currently supporting Java versions up to 13. 
For incompatible Java versions, \textit{tsDetect} may encounter syntax parsing errors during detection, which could prevent the identification of test smells, thus impacting the effectiveness of smell elimination.
Given that test smell detection tools are continually evolving, \appname is designed to be highly extensible, allowing for the replacement of the built-in test smell detection tool. 
When encountering incompatible Java versions, a more robust detection tool can be integrated to address undetected smells, thereby improving the overall performance of the test refactoring process.

\section{RELATED WORK}
\label{sec:relatedwork}
\noindent \textbf{The Impact and Refactoring of Test Smells.}
Several studies~\cite{soares2020refactoring,pizzini2023sentinel,pizzini2022behavior,nagy2022co,damasceno2022analyzing,soares2022refactoring,kashiwa2021does,santana2024empirical,martins2024comprehensive,damasceno2023test} have focused on the impact of test smells, including how they affect the development process, software maintenance, and comprehension. 
To investigate how much developers acknowledge the presence of test smells, Soares et al.~\cite{soares2020refactoring} conducted a study with 73 experienced open-source developers across 272 projects. 
They analyzed preferences and motivations related to 10 identified test smells by comparing the original and refactored versions of test code. 
The results showed that 78\% of developers acknowledged the negative impact of test smells and preferred refactored tests.
Besides, they explored the use of new JUnit 5 features to eliminate and prevent test smells in another study~\cite{soares2022refactoring}. 
In an empirical study on 485 popular Java open-source projects from GitHub, Soares et al. found that only 15.9\% of projects used the JUnit 5 library. 
By applying seven JUnit 5 features to address test smells, they conducted a survey of 212 developers and submitted 38 pull requests, achieving a 94\% acceptance rate among respondents.

\noindent \textbf{Test Smell Detection.}
There are many research methods and tools~\cite{peruma2020tsdetect,yang2024lost,aljedaani2021test,palomba2018automatic,pontillo2024machine,wang2021pynose,danphitsanuphan2012code,fernandes2022tempy,bodea2022pytest,virginio2020jnose} available for detecting test smells across different programming languages.
Palomba et al.~\cite{palomba2018automatic} developed \textit{TASTE}, an automated textual-based tool for detecting several types of test smells. 
Compared to previous structure-based detection methods, this tool improved smell detection effectiveness by 44\%.
Peruma et al.~\cite{peruma2020tsdetect} recently developed a tool called \textit{tsDetect}, capable of detecting 19 test smells in Java. 
It used a set of detection rules to locate existing test smells in test code.
Wang et al.~\cite{wang2021pynose} proposed a tool called \textit{PYNOSE}, designed to detect 17 types of test smells within Python's standard \textit{Unittest} framework, and introduced a new test smell type called \textit{Suboptimal Assert}.
\textit{PYNOSE} was available as a plugin for \textit{PyCharm}, and in an empirical study conducted across 248 Python projects, they found that 98\% of projects contained at least one type of test smell.
Unlike work focused on test smell detection, our research aims to explore how the latest large language models could improve the performance of eliminating detected smells.


\section{Conclusion and Future Work}
\label{sec:conclusion}
This paper introduces \appname, a tool that focuses on the automatic detection and refactoring of test smells.
In this work, we explore how leveraging refactoring DSLs and incorporating test context knowledge optimize and enhance the ability of open-source LLMs to automatically refactor test code. 
\appname supports multiple test smell types and eliminates smells at varying granularities.
Compared to existing methods, \appname demonstrates greater effectiveness in eliminating various types of test smells from code.
In the future, we plan to extend our refactoring rules to support additional programming languages, such as Python and C/C++. 
Additionally, we intend to investigate the impact of using different LLM sizes and parameters on the test refactoring process.

\section{Data Availability}
Our tool is available at~\cite{utrefactor}. 

\begin{acks}
This work was supported by the National Key R\&D Program of China (No. 2024YFB4506400) and sponsored by CCF-Huawei Populus Grove Fund.
\end{acks}

\balance
\bibliographystyle{ACM-Reference-Format}
\bibliography{main}
\end{document}